\RequirePackage{fix-cm}
\makeatletter
\def\cl@chapter{}
\def\makeheadbox{}
\makeatother

\documentclass[twocolumn]{svjour3} 
\smartqed
\def\makeheadbox{}

\usepackage[square,numbers]{natbib}
\usepackage{url}
\usepackage{graphicx}
\usepackage{amsmath}
\usepackage{upgreek}
\usepackage{parskip}
\usepackage[locale=UK,
number-mode=math,
unit-mode=text,
per-mode=symbol,
number-unit-product=\ ,
retain-zero-exponent=false]{siunitx}
\usepackage{booktabs}
\usepackage{paralist, todonotes}

\usepackage[hidelinks]{hyperref}
\usepackage{cleveref}

\newcommand{\kindex}[2]{\ensuremath{{#1}_{\scalebox{0.5}{#2}}}}
\def\Rey{\mbox{\it Re}}   

\usepackage{etoolbox} 
\robustify{\kindex}

\graphicspath{{../figurematter/plots/}{../figurematter/latex/figs/}}

\begin{document}
\title{On the parametrisation of motion kinematics for experimental aerodynamic optimisation
}
\author{Christoph Busch \and Alexander Gehrke \and Karen Mulleners}
\institute{
\at
Unsteady Flow Diagnostics Laboratory (UNFoLD)\\
Institute of Mechanical Engineering (IGM)\\
\'Ecole Polytechnique F\'ed\'erale de Lausanne (EPFL)\\
Lausanne, Switzerland\\
\email{karen.mulleners@epfl.ch}
}

\titlerunning{Parametrisation of motion kinematics}

\date{Received: date / Accepted: date}
\maketitle

\begin{abstract}
The levels of agility and flight or swimming performance demonstrated by insects, birds, fish, and even some aquatic invertebrates, are often vastly superior to what even the most advanced human-engineered vehicles operating in the same regimes are capable of.
Key to this superior locomotion is the animal’s manipulation of the generation and shedding of vortices through optimal control of their motion kinematics.
Many research efforts related to biological and bio-inspired propulsion focus on understanding the influence of the motion kinematics on the propulsion performance and on optimising the kinematics to improve efficiency or manoeuvrability.
One of the first challenges to tackle when conducting a numerical or experimental optimisation of motion kinematics of objects moving through a fluid is the parameterisation of the motion kinematics.
In this paper, we present three different approaches to parameterise kinematics, using a set of control points that are connected by a spline interpolation, a finite Fourier series, and a reduced order modal reconstruction based on a proper orthogonal decomposition of a set of random walk trajectories.
We compare the results and performance of the different parameterisations for the example of an experimental multi-objective optimisation of the pitching kinematics of a robotic flapping wing device.
The optimisation was conducted using a genetic algorithm with the objective to maximise stroke average lift and efficiency.
The performance is evaluated with regard to the diversity of the randomly created initial populations, the convergence behaviour of the optimisation, and the final Pareto fronts with their corresponding fitness values.
The suggested approaches perform equally well and yield fitness values that are in close proximity for the different kinematic functions and different number of input parameters.
The main differences are concerned with the implementation of experimental constraints and minor variations in the shape of the Pareto-optimal motions are observed.
Dedicated applications for each approach are suggested.
\end{abstract}
\maketitle


\section{Introduction}
Unsteady locomotion of flying animals such as birds and insects have inspired the development of micro air vehicles, which are less than \SI{15}{\centi\meter} in size and operate in a Reynolds number range from \numrange{10}{10000} \cite{shyy_flapping_1999,shyy_aerodynamics_2011,shyy_recent_2010}.
Their field of application includes information gathering in confined spaces, areal mapping, or the transport of small goods \cite{floreano_science_2015,taylor_experimental_2010}.
Flapping wing configurations are a suitable manner of propulsion for micro air vehicles, since they offer better efficiency at low Reynolds number ($\Rey < \num{100}$) and an improved manoeuvrability in comparison to fixed and rotating wing configurations \cite{bayiz_hovering_2018,hawkes_fruit_2016}.
The development of complex flapping wing kinematics for various flight scenarios poses a challenge in this field.
A similar challenge is the selection of optimal motion kinematics of other bio-inspired propulsion systems such as underwater vehicles mimicking fish propulsion.
A standardised optimisation procedure facilitates the identification of motions with desired characteristics.
The motion kinematics have to be parameterised by a finite number of variable parameters prior to optimisation to reduce the dimensionality of the solution space and allow for optimal solutions to be found in an affordable way within a reasonable timeframe.
The optimisation solution landscape can dependent strongly on the parameterisation and possible optima might be missed \cite{mandre_optimizing_2019}.
The ideal parameterisation has as little variable parameters as necessary to reduce the computational complexity of the optimisation problem without restricting the solution space or creating a bias in a certain direction.
Yet, any parameterisation will exclude possible solutions due to discretisation in comparison to a continuous solution in an infinite-dimensional solution space and it is vital to carefully select the most suitable approach for each optimisation problem at hand.

Past efforts to optimise bio-inspired motion kinematics initially focussed on simple linear and harmonic motions.
\citet{tuncer_optimization_2005} numerically optimised a flapping airfoil for combined maximum thrust and efficiency.
They parameterised the sinusoidal plunge and pitching motions with a variable amplitude and phase shift for a fixed frequency.
\citet{berman_energy-minimizing_2007} implemented a quasi-steady model of insects with a hybrid optimisation algorithm.
Their kinematic function enables a continuous transition of the pitching angle evolution $\beta$ with the parameter $C$ between sinusoidal and trapezoidal motions according to
\begin{equation}
    \beta(t) = \frac{\kindex{\beta}{m}}{\tanh(C)} \tanh\left(C \sin(2\pi f t + \Phi)\right) + \kindex{\beta}{0}
\end{equation}
with $f$ being the flapping frequency, $\Phi$ the phase shift between the stroke and pitching motions, $\kindex{\beta}{0}$ the offset angle and a scaling factor $\kindex{\beta}{m}$ for further modulations.
The temporal evolutions of the stroke and elevation angles were defined similarly, yielding a total of twelve variable parameters.
Solutions for these twelve parameters were determined that maximised the aerodynamic efficiency while providing enough lift to support the body weight of different insects.
The aerodynamic performance can be further increased by more complex, non-harmonic, and asymmetric kinematics, which have to be defined using more parameters or different base functions.
The pitching kinematics of insects in nature are often asymmetric which could be an evolutionary adaption to improve the performance \cite{liu_size_2009}.
\citet{martin_experimental_2018} were among the first to implement a kinematic motion function with the possibility to represent asymmetric motions.
\citet{liu_size_2009} performed an optimisation of the motion of a pectoral fish fin which was defined as a complex combination of trigonometric expressions for all three spatial angles.
The motion kinematics were successfully optimised for a combined objective of minimal side-thrust and maximal efficiency.
They noticed an increasing difficulty in finding the global optimum with increasing number of parameters describing the kinematics.
\citet{mandre_optimizing_2019} investigated the influence of the parameterisation with a finite Fourier series on a heaving and pitching hydrofoil.
Small modifications of the parameterisation can lead to a significantly altered solution landscape with a bias towards small but pronounced local optima.
\citet{gehrke_phenomenology_2021} studied pitching motions with increased complexity, which are able to mimic more closely the motions observed in nature.
The stroke motion was fixed and the elevation motion was omitted.
The pitching angle was defined by a set of control points in the angle-time domain, which were connected with a third-degree spline.
The resulting optimal pitching kinematics forming the Pareto front in the stroke average lift versus efficiency diagram vary significantly and highlight the potential of asymmetric and more complex kinematics to push the performance envelope of bio-inspired micro aerial vehicles.

Here, we will present three different approaches to define and parameterise kinematics for optimisation studies.
The different approaches will be presented, compared, and evaluated for the example of the experimental optimisation of the pitching kinematics of a flapping wing, building upon the work presented in \citet{gehrke_genetic_2018} and \citet{gehrke_phenomenology_2021}.
An optimisation is performed for each kinematic function with $\num{6},\num{12}$ and $\num{18}$ parameters, leading to more than \num{30000} individual experiments in total.
The different approaches to parameterise motion kinematics for optimisation selected here are:
\begin{itemize}
\item control points connected by a spline interpolation,
\item a finite Fourier series, and
\item a linear combination of modes determined by a modal decomposition of kinematics created by a random walk
\end{itemize}

\section{Materials and Methods}
\subsection{Experimental set-up}
\begin{figure}
\centering
\includegraphics[width=\linewidth]{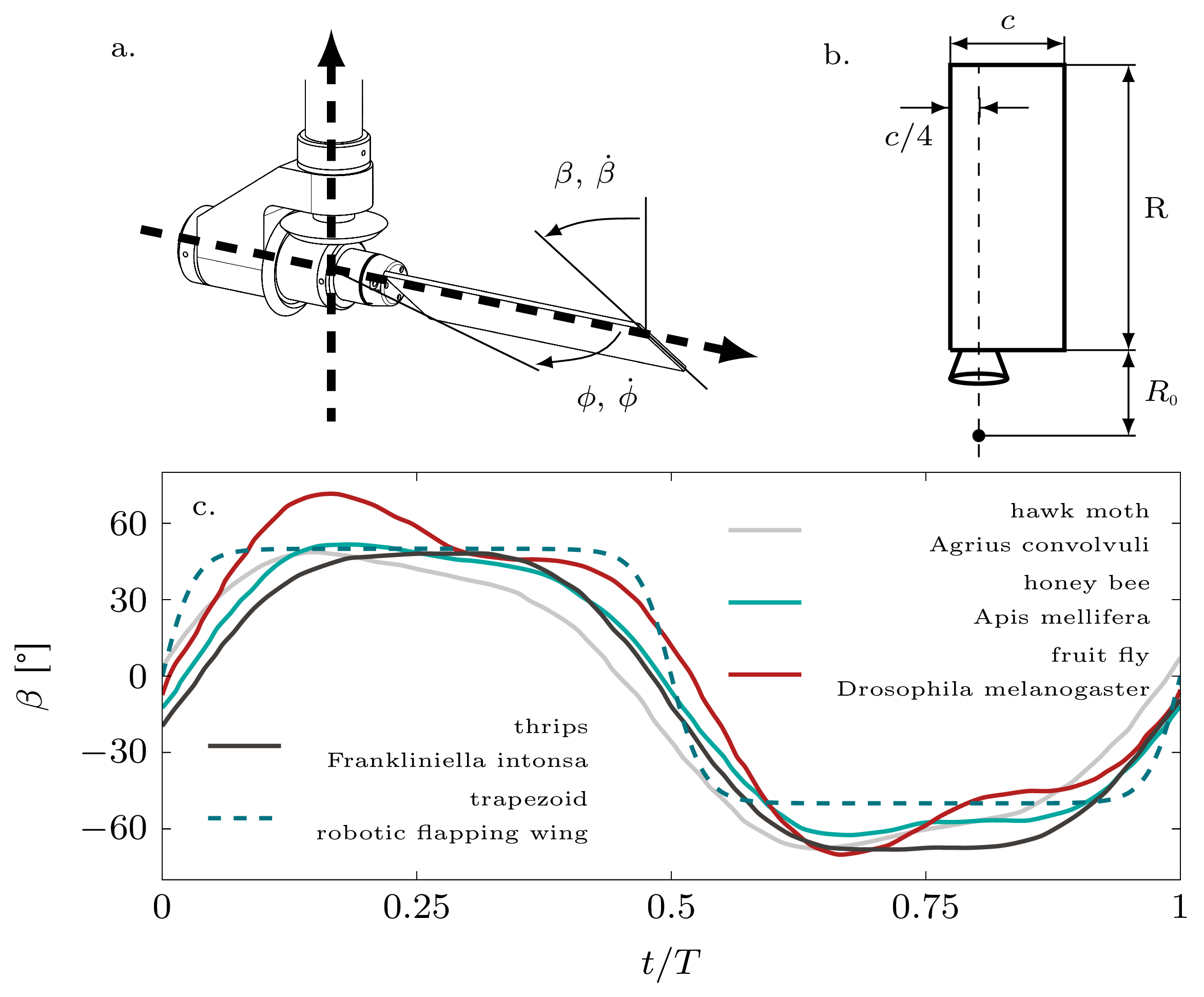}
\caption{a. Flapping wing mechanism with stroke angle $\phi$ and pitch angle $\beta$ for the hovering motion, b. wing geometry and axis of rotation, c. pitch angle $\beta$ kinematics of various natural fliers, adopted from~\cite{liu_size_2009} and trapezoidal pitch angle kinematics from a robotic flapper~\cite{Bhat.2020}.}
\label{fig:experimental_setup}
\end{figure}

The kinematics for the different optimisation studies are evaluated with a robotic flapping wing mechanism immersed in an octagonal water tank (see also~\cite{gehrke_genetic_2018, gehrke_phenomenology_2021}).
The wing is a rectangular flat plate with a chord of $c = \SI{34}{\milli\metre}$ and a span of $R = \SI{107}{\milli\metre}$ (\cref{fig:experimental_setup}).
The flapping frequency of the system is $f = \SI{0.25}{\hertz}$ with a peak-to-peak stroke amplitude of $\phi = \ang{180}$.
The flapping frequency and stroke amplitude remain constant throughout this study and lead to a Reynolds number of $Re = 2 \phi f c \kindex{R}{2} / \nu = 4895$ at the radius of the second moment of area $\kindex{R}{2}$, for a kinematic viscosity of $\kindex{\nu}{\SI{20}{\celsius}} = \SI{1.00e-6}{\meter\squared\per\second}$ for all experiments~\cite{gehrke_phenomenology_2021, sum_wu_scaling_2019}.
These characteristic parameters are selected to match typical flapping wing vehicles and larger insects such as the hawk moth~\cite{liu_size_2009, shyy_recent_2010}.
The wing is actuated by two servo motors (Maxon motors, type RE35, \SI{90}{\watt}, \SI{100}{\newton\milli\meter} torque, Switzerland) at the top of a 2-axis shaft system which controls the pitch and stroke angles of the wing (\cref{fig:experimental_setup}).
The wing moves for a total tip-to-tip amplitude of \SI{0.47}{\metre} which results in a \SI{6.91}{c} tip clearance to the tank for an outer diameter of \SI{0.75}{\metre}.
Previous studies have confirmed that no wall effects are present for a tip clearance larger than \SI{5}{c}~\cite{manar_tip_2014, krishna_flowfield_2018}.
The flapping wing apparatus was specifically designed to run unsupervised for large amounts of time and to execute complex motion profiles with high repeatability.
Preliminary tests have determined the maximum error between prescribed motion and motor response on the mechanism to be $< \ang{0.1}$ for high acceleration kinematics~\cite{gehrke_phenomenology_2021}.
The motors are controlled by a motion control board (DMC-4040, Galil Motion Control, USA).
A six-axis force and torque transducer (Nano17, ATI Industrial Automation, USA) is installed at the wing root in \cref{fig:experimental_setup}b to capture the time resolved aerodynamic loads.
The sensor is calibrated with \textit{SI-12-0.12} to record at a resolution of \SI{3.13}{\milli\newton} for force and \SI{0.0156}{\newton\milli\meter} for torque measurements.
A data acquisition card (National Instruments, USA) is used to record the force and torque transducer signal at a sample frequency of \SI{1000}{\hertz}.
The lift and power data was filtered with a zero phase delay low-pass \num{5}th order digital Butterworth filter for the phase-averaged, time-resolved plots.
The filter's cut-off frequency was selected to be \num{12} times the hovering frequency of the system.

\subsection{Kinematic functions}
Three different approaches to parameterise motion kinematics are selected here:
\begin{inparaenum}[a)]
\item control points connected by a spline interpolation,
\item a finite Fourier series, and
\item a linear combination of modes determined by a modal decomposition of kinematics created by a random walk.
\end{inparaenum}
The different parameterisation were selected by their ability to mimic trapezoidal and sinusoidal pitch angle profiles commonly seen on robotic flapping wing devices as well as more complex pitch angle kinematics observed on nature's fliers (\cref{fig:experimental_setup}c).
We created kinematic functions using the three approaches with three different parameter counts ($p = 6,12, 18$).
The first $(p-1)$ parameters are used to modify the pitching angle evolution and the phase shift is controlled with the last parameter $\kindex{\Delta t}{0}$ for all kinematic functions.
The motions are described by the pitching angle $\beta$ (\cref{fig:experimental_setup}).
The sinusoidal stroke angle $\phi$ remains unchanged throughout the optimisations and the elevation angle is kept at zero due to its negligible influence on the lift production \cite{liu_size_2009}.

\subsubsection{Control points connected by splines}

\begin{figure*}[tb!]
\centering
\includegraphics[width=\linewidth]{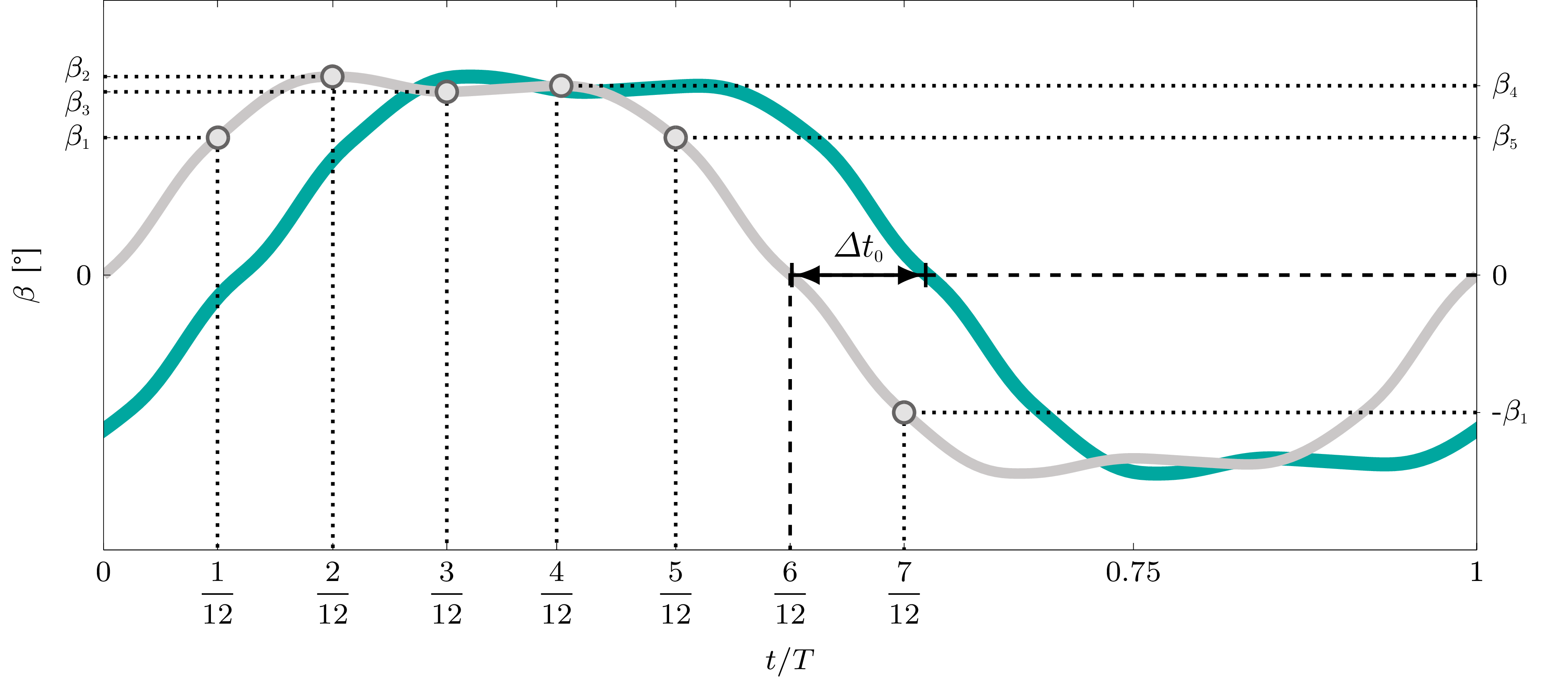}
\caption{Example of the parameterisation of the pitch angle kinematics using control points connected by a spline interpolation with $p=6$ parameters.
Five parameters describe the pitch angles ($\kindex{\beta}{1}, ..., \kindex{\beta}{5}$) of the five control points that are equidistantly distributed within a half stroke.
A fifth degree spline connects the five control points and the half half stroke profile is point mirrored to create the full symmetric pitch angle profile in grey.
The sixth parameter $\kindex{\Delta t}{0}$ introduced a phase shift leading to the final pitching profile in colour.}
\label{fig:pchip_definition}
\end{figure*}

As a first intuitive approach, we select a limited number of control points that will be connected using spline interpolation.
To start, we create a symmetric pitch profile by distributing $(p-1)$ control points equidistantly throughout a half-stroke which represents the most efficient use of the available parameters.
Freely spaced control points could lead to large gradients for higher parameter counts which can not be executed by our experimental set-up.
In a way, the temporal spacing of the control points acts as a build-in low-pass filter for the pitch angle gradients.
If the phase of the control points is not fixed, the phase bounds of the individual points depend on each other which makes the selection of the kinematics slightly more complicated.
Depending on the specific problem at hand, it is possible to place more control points in certain parts of the motion where larger variations are required to optimise and reduce the parameter count.
The beginning and end of the half-stroke are fixed at $\beta=\ang{0}$ and the control points are solely specified by their pitch angle \kindex{\beta}{n}, with $1\leq n\leq(p-1)$ as indicated in \cref{fig:pchip_definition}.
Next, the control points are connected by a fifth degree spline interpolation to minimise the local acceleration $\ddot{\beta}$.
By point-mirroring the first half-stroke around $(t/T,\beta)=(0.5,0)$, we obtain a symmetric pitching profile for the back and forth stroke, represented by the grey curve in \cref{fig:pchip_definition}.
Finally, we added a phase-shifted \kindex{\Delta t}{0} to allow for advanced and delayed wing rotation with respect to the stroke reversal, leading to the coloured curve in \cref{fig:pchip_definition}.
This procedure allows us to created complex non-linear and non-harmonic motions with high curvatures.
The main advantage of the control point parameterisation, is its intuitive and direct local control of the pitching profiles.
Local adjustments can be made by moving individual control points and the values of the control points are directly linked to the pitch angle values, which facilitates the formulation of pitch angle constraints due to mechanical and motor limitations.
The distribution of the control points along the time axis can be easily varied to adapt the approach to specific restrictions or requirements for a broad range of applications.
On the downside, we noticed that the creation of a random set of control point splines is computationally demanding and required $\SI{22.8}{\second}$ on average for motions with $12$ parameters on a typical desktop computer.
The computational time increases exponentially with the parameter count and takes a couple of minutes for the tests with $18$ parameters.

\subsubsection{Fourier-series kinematics}

\begin{figure*}[tb]
\centering
\includegraphics[width=\linewidth]{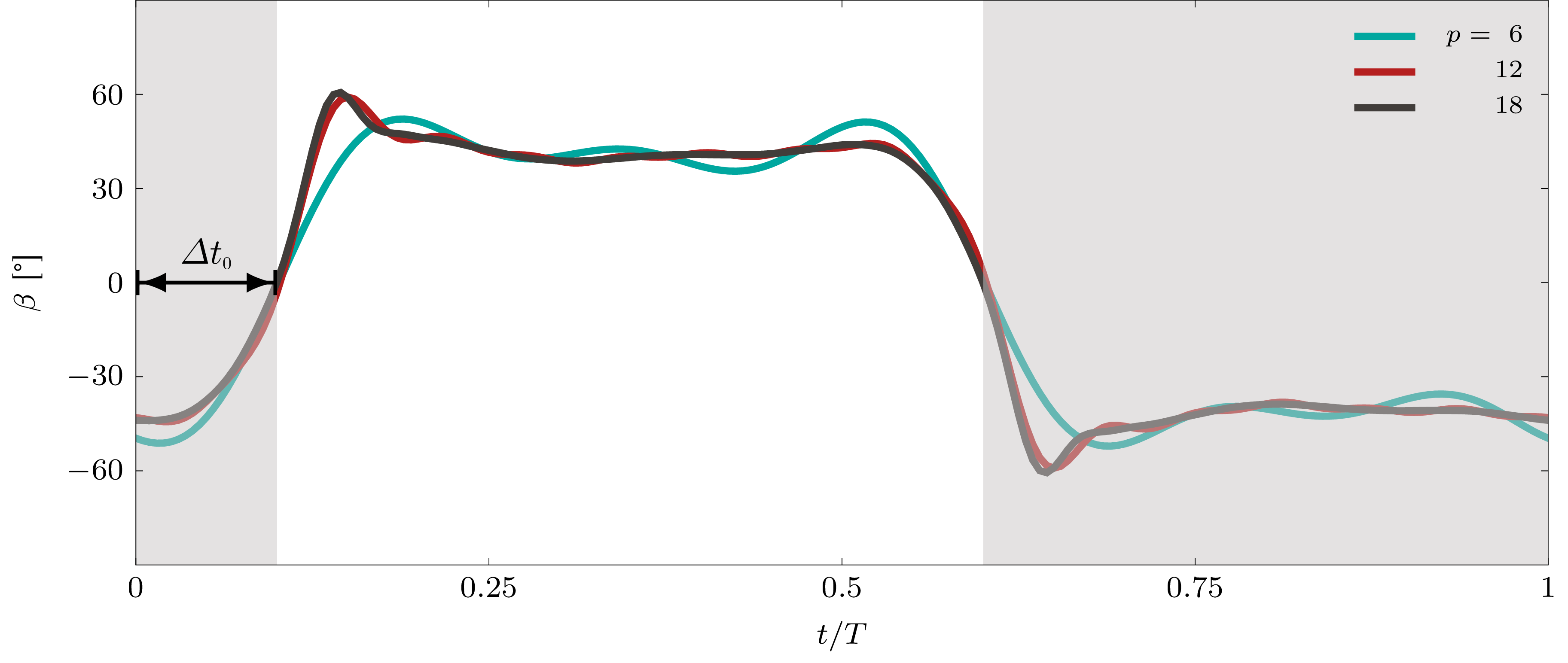}
\caption{Example of the parameterisation of the pitch angle kinematics using a finite Fouries series $p=6$, $12$, and $18$ parameters.
The last parameter $\kindex{\Delta t}{0}$ introduced the phase shift.}
\label{fig:fourier_kin}
\end{figure*}

In the second approach, we use a finite Fourier sine-cosine series to define our pitching kinematics:
\begin{equation}
\beta\left( t \right) = \sum\limits_{ k=1,3,5,...}^{N}\Big( \kindex{a}{k} \cos\left( 2\pi k\kindex{f}{0} t \right) + \kindex{b}{k} \sin\left(2\pi k\kindex{f}{0} t \right) \Big)
\label{eq.fourier}
\end{equation}
with $\kindex{a}{k}$ the cosine coefficients, $\kindex{b}{k}$ the sine coefficients, and $\kindex{f}{0}$ the flapping frequency.
For the current application, we use only the odd coefficients.
The first odd sine-coefficients $\kindex{b}{k}$ are responsible for the base oscillation whereas the odd cosine-coefficients $\kindex{a}{k}$ are used for smaller modulations.
The ability to represent higher curvatures, as they appear in trapezoidal motions, mainly depends on the number of Fourier terms included.
A finite Fourier series is not ideal to describe sudden jumps or discontinuities and tends to display an overshoot followed by decaying oscillations near sharp gradients.
This behaviour is known as the Gibbs phenomenon and can be reduced by adding a Lanczos-$\sigma$-factor \cite{hamming_numerical_1986}.
The Lanczos-$\sigma$-factor is defined as:
\begin{equation}
\sigma\left( \frac{ k }{ m } \right) = \frac{ \sin\left( \frac{ k }{ m } \pi\right) }{ \frac{ k}{ m }\pi  }
\label{eq:lanczos-sigma-factor}
\end{equation}
with $m$ as the last plus one summing index of a finite Fourier series.
The Fourier series based pitching kinematics used in this paper are described by a finite Fourier series with Lanczos-$\sigma$ correction factor:
\begin{equation}
\begin{split}
\beta\left( t \right) = \sum\limits_{k=1,3,5,...}^{ (p-1)/2 } & \sigma  \left( \frac{ 2k }{p+1} \right) \\ & \Big( \kindex{a}{k} \cos\left( 2\pi k\kindex{f}{0} t \right)  + \kindex{b}{k} \sin\left(2\pi k\kindex{f}{0} t \right)\Big)
\end{split}
\label{eq:fourier_with_sigma}
\end{equation}
with $(p-1)$ parameters, $(p-1)/2$ cosines and $(p-1)/2$ sine coefficients.
The obtained motion kinematics are periodic, continuous in all derivatives and can easily be expressed in a closed form.
The half strokes can be asymmetric around the quarter stroke.
The phase-shift parameter \kindex{\Delta t}{0} is again applied at the end to create advanced or delayed wing rotations with respect to the stroke reversal.
Exemplary motions created using $6$, $12$, and $18$ parameters are presented in \cref{fig:fourier_kin}.

\subsubsection{Modal reconstruction}

\begin{figure*}
\centering
\includegraphics[width=\linewidth]{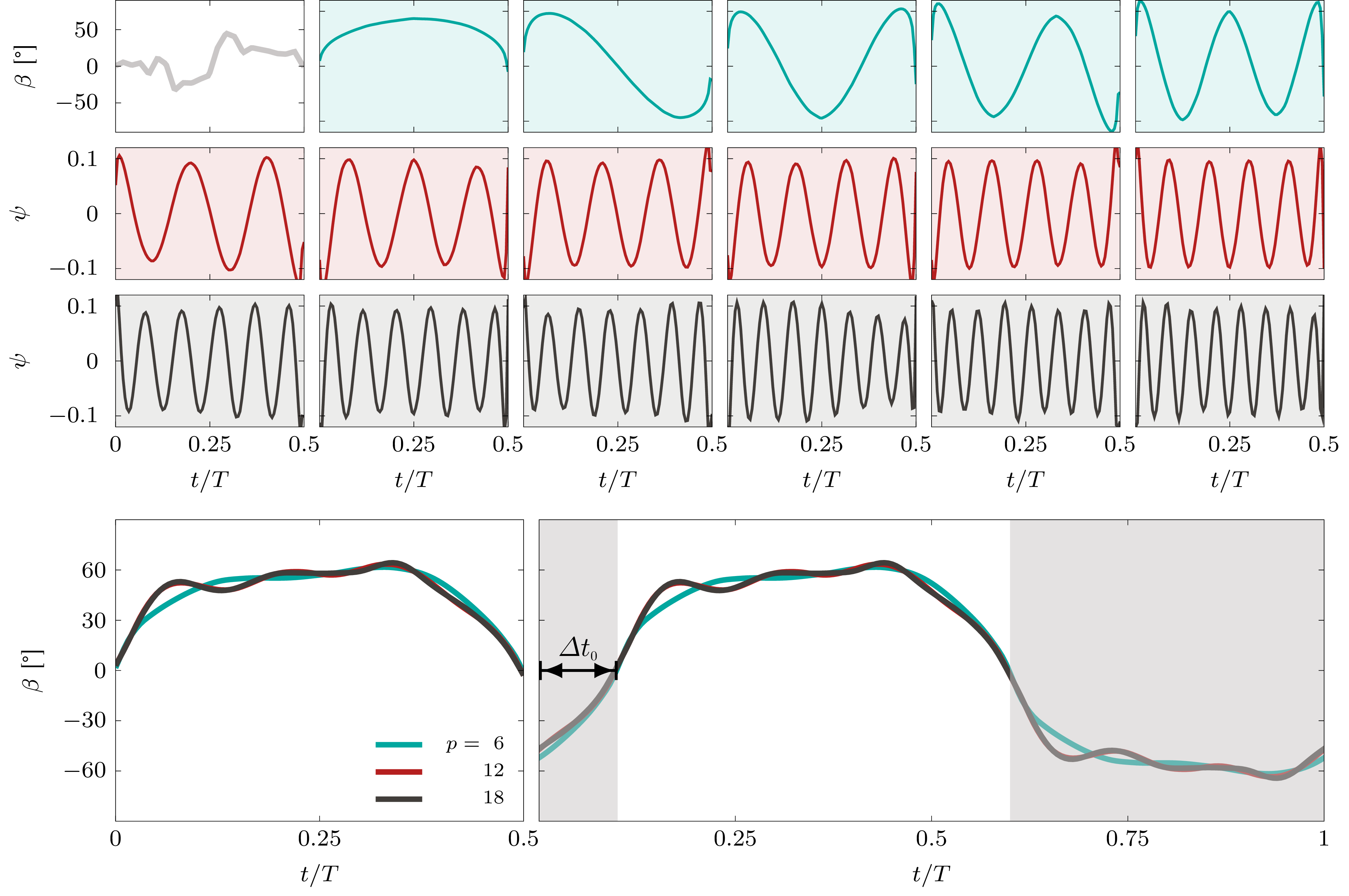}
\caption{Example of an individual random walk-based motion (top left) and the first $17$ eigenmodes of the proper orthogonal decomposition of a family of $2 \times 10^5$ random walk-based motions.
The first $5$ modes are presented with a green background, modes $6$ to $11$ with a red background, and modes $12$ to $17$ with a grey background.
Examples of new motions created as a linear combination of the first $5$, $11$, and $17$ modes are presented in the bottom left.
The half stroke motions are point-mirrored around $(T/2,0)$ to obtain a full stroke and the phase shift is applied to create the final motions in the bottom right.
}
\label{fig:modes}
\end{figure*}

In the third approach, we use a low-order eigenmode reconstruction.
The eigenmodes are obtained by a modal decomposition of a family of available arbitrary kinematics.
This parameterisation approach is particularly elegant if a family of kinematics is known.
For example, if a database of kinematics is available from direct observations of insects or fish motions, we can first reduce the dimensionality of the available kinematics using modal decomposition.
Measured and new kinematics can then be created by low-order reconstruction of the most dominant kinematic modes.
Here, we do not have a database of measured kinematics at our disposal.
To demonstrate the concept of eigenmode reconstruction as a parameterisation approach, we have artificially generated a family of randomised kinematics based on a biased random walk algorithm.
The random walk motion generator is not an essential part of the approach it merely serves as bypass to obtain a generalised motion database.
The random walk algorithm takes randomly sized steps in a discretised half-stroke.
The random expectation value distribution for the individual step sizes is defined such that their sum equals zero over one half-stroke.
Each step is defined as the expectation value plus a random fluctuation.
The motions starts at $\beta=\ang{0}$ and is set again to zero at the end of the half-stroke.
This creates a wider spread of values and higher gradients at the end of the half-stroke in comparison to the start of the half-stroke.
This bias is resolved by line-mirroring motions at mid-half-stroke ($T/4$). 
An example of a random walk-based motion is presented in the top left panel of \cref{fig:modes}.
A large amount of motions ($\mathcal{O}(\num{e5})$) is created within a few minutes with this method.

The family of random walk-based motions are then decomposed using proper orthogonal modal decomposition (POD).
All modes start and end with a zero passage and are nearly symmetric (\cref{fig:modes}).
The symmetry is expected to improve with a larger data basis of random kinematics whereas the computational complexity of the POD is the limiting factor for the number of input motions.
The modes exhibit high similarity with Legendre polynomials starting from the second polynomial.
The POD ensures the optimality of the projection space for a given set of training motions and the optimal use of parameters.
A large variety of complex, asymmetric motions and oscillations can be defined by a linear combination of the first $(p-1)$ eigenmodes:
\begin{equation}
\beta(t)= \sum\limits_{n=1}^{p-1} \kindex{a}{n} \sqrt{2\kindex{\lambda}{n}} \,\kindex{\psi}{n}(t)
\end{equation}
with \kindex{\psi}{n} the POD eigenmodes, \kindex{\lambda}{n} the corresponding eigenvalues, and \kindex{a}{n} the parameterisation coefficients.
The first $17$ eigenmodes, \kindex{\psi}{1}-\kindex{\psi}{17}, are presented in \cref{fig:modes} for the proper orthogonal decomposition of the family of random walk based motions used in this paper.
The coefficients \kindex{a}{n} are considered normalised and have values confined between \num{-1} and \num{-1}.
The motions are point-mirrored around $(T/2,0)$ to obtain a full stroke and the phase shift is applied to create the final motion as depicted in the bottom row of \cref{fig:modes}.
Discontinuities in the gradient between the end and beginning of each half-stroke cannot be executed by the motors and are mitigated by a robust local regression smoothing, which is tuned to mainly affect the sections around stroke reversal.
Discontinuities in the velocity $\dot{\beta}$ appear for randomly created kinematics in the early populations and diminish when the convergence progresses.

\subsection{Genetic algorithm optimisation}

The pitching kinematics of our robotic flapping wing device have been optimised using the multi-objective genetic algorithm optimisation algorithm (\texttt{gamultiobj}) from the global optimisation toolbox of Matlab\textsuperscript{\tiny\textregistered} \cite{matlab_global_2020}.
A genetic algorithm is a meta-heuristic optimisation method, suited to find the global optimum in a non-linear solution space with a large number of degrees of freedom.
It mimics the process of natural selection known from evolution by recombination and mutation.
An initial population of $200$ individuals, with a uniform distribution of the parameters within their bounds is created to start the optimisation.
Each individual corresponds to one periodic flapping motion.
The fitness values for two optimisation objectives are directly measured during experiments for each motion.
The optimisation objectives are the maximum stroke average lift $\kindex{\overline{C}}{L}$ and the maximum stroke average efficiency $\overline{\eta}$.
The stroke average efficiency $\overline{\eta}$ is defined here as:
\begin{equation}
\overline{\eta} = \frac{\kindex{\overline{C}}{L}}{\kindex{\overline{C}}{P}}.
\end{equation}
The overline indicates stroke average quantities.
The lift and power coefficients,\kindex{C}{L} and \kindex{C}{P}, are defined as
\begin{equation}
\kindex{C}{L} = \frac{ L }{ \frac{ 1 }{ 2 } \rho R c \overline{U}^{2}}\quad, \quad
\kindex{C}{P} = \frac{ P }{ \frac{ 1 }{ 2 } \rho R c \overline{U}^{3}}\quad.
\end{equation}
The dimensional lift $L$ and power $P$ are normalised with the density $\rho$, the wing span $R$, the chord length $c$ and the stroke average velocity $\overline{U}$.
Each motion is executed over eight cycles.
The stroke average quantities are determined based on the last four cycles to exclude any influence of start-up transients.
At the selected Reynolds number, the load responses are extremely repeatable and the difference between the individual cycle averaged lift coefficients and the average coefficient over the last four cycles is below \SI{5}{\percent} after the first cycle and below \SI{2}{\percent} after three cycles.
Variations up to \SI{20}{\percent} in the lift coefficient are observed in the first cycle for the most efficient kinematics.
The evaluations of the individual experiments take $\SI{39}{\second}$ each, including a settling time for the water in the tank.
The fittest individuals produce offspring by recombination and mutation.
Recombined individuals account for \SI{60}{\percent} of the following population with the parameters calculated by a randomised linear combination $\textit{child} = \textit{parent 1} + R ( \textit{parent 2} - \textit{parent 1})$, with $R$ a randomly picked number from a  uniform distribution covering the interval $[0,1]$ \cite{matlab_global_2020}.
The remaining \SI{40}{\percent} of the new population are created by random mutation of the individuals' parameters.
No clones from the elite are transferred into the next generation.
The fitness values of the new individuals are determined and the procedure is repeated until convergence is detected, after which the optimisation is stopped.
The final output of the genetic algorithm optimisation is a global Pareto front, which consists of all non-inferior solutions in the $\kindex{\overline{C}}{L}$ versus $\overline{\eta}$ space.
The solutions on the Pareto front are called non-inferior solutions if the value for each objective can only be improved by decreasing the value of another one.
The results of $9$ optimisations will be presented and compared here.
We have considered each of the parameterisation approaches with $p=6$, $12$, and $18$ parameters.

The experimental set-up is subjected to mechanical and safety constraints to protect the equipment.
The maximum pitching angle is limited by the range of free movement in the experimental set up such that $|\kindex{\beta}{max}| \leq \SI{90}{\degree}$.
The minimum angle of $\kindex{\beta}{min} = \ang{0}$ is imposed for the parameterisation of the initial half stroke before applying the phase-shift $\kindex{\Delta t}{0}$ to avoid ambiguous solutions where advance or delayed rotation is not governed entirely by the phase-shift.
The phase shift is bound by $[-\pi/2, \pi/2]$.
The minimum and maximum pitch angle bounds are easy to define for the control point approach by directly limiting the values of $\beta$ of the control points.
This does not apply for the Fourier series and modal reconstruction based parameterisations.
Here, the extreme angles are the result of a combination of different parameters.
We opted here to keep the allowed parameter ranges as wide as possible and to reject proposed kinematics that are not safe or not feasible for the experimental mechanism before execution.

The most important constraint for our robotic flapper is the pitch angle acceleration.
High accelerations put stress on the load cell due to the inertial forces on the wing and have to be limited to avoid damage.
The maximally allowed pitch angle accelerations are $|\kindex{\ddot{\beta}}{max}| \leq \SI{1800}{\degree\per\second\squared}$.
The majority of randomly created individuals for the control point and modal reconstruction approach exceed the acceleration constraint.
This leads to a validity rate of $ \leq \SI{10}{\percent}$ in combination with the constraint $|\kindex{\beta}{max}| \leq \SI{90}{\degree}$.
The parameter bounds are tightened for the Fourier series approach to keep the share of invalid motions within a range of \SIrange{10}{15}{\percent}.
A simple restriction of the parameter space does not have the same effect for the splines connecting control points and the modal reconstruction.
Here, a pre-selection of kinematics is necessary to obtain a full initial population of $200$ individuals.
The average and peak acceleration values of the randomly created motions increase with the parameter count and so does the share of randomly generated motions that cannot be executed.
The pre-selection process takes a couple minutes for the modal reconstruction with $6$ parameters and up to $\SI{4}{\hour}$ for the control points connected by splines with $12$ parameters on a standard desktop computer.
The control point approach with $18$ parameters has very tightly spaced control points, which lead to high local accelerations above the limit value but they only appear for a very short time such that their risk of damaging the set-up is deemed low.
The constraint of a minimum angle $\kindex{\beta}{min}\geq \SI{0}{\degree}$ in combination with the acceleration and velocity constraints is too strict for the modal reconstruction approach to create an initial population in a reasonable amount of time.
By loosening the lower angular constraint $\kindex{\beta}{min}$, and allowing $\kindex{\beta}{min}<0$, we obtain more kinematics that have a maximum acceleration below the mechanically allowed limit.
All parameterisation approaches required a customised implementation of the constraints and adapted selection of the parameter bounds.
The implementation of the experimental constraints was most straight forward for the control point based approach due to the more direct local access to the kinematics.

\subsection{Convergence criterion}
To monitor the progress of the optimisations and to decide when to consider the result to be converged, we have implemented a generational distance measure.
The generational distance allows for a quantitative comparison between the optimisations of the different kinematic functions possible.
It is a common measure to determine optimisation progress in genetic algorithms if no optimal Pareto front as reference is known \cite{audet_performance_2020}.
The generational distance (GD) is defined by \citet{veldhuizen_multiobjective_1999} as:
\begin{equation}\label{eq:gd}
\textrm{GD}\left( S,P \right)= \frac{ 1 }{ \left\lvert S \right\rvert } \Bigg( \sum\limits_{ s\in S }{ \,\min\limits_{r\in P}\parallel F \left( s \right)-F \left( r \right)\parallel^{ p }} \Bigg)^{ \frac{ 1 }{ p } },
\end{equation}
with $S$ the set of Pareto front members of the current generation, $P$ the Pareto set members of the previous generation, $F$ the objective function, and $p$ the selected distance norm.
The generational distance describes the average distance the Pareto front members are shifted between successive generations and represents the optimisation progress.
The individual distances are measured between the point of the current front and the closest point from the previous front.
All distance measures for the efficiency $\overline{\eta}$ and stroke over lift coefficient $\kindex{\overline{C}}{L}$ are normalised by the average range of values for the first ten generations (or all generations if less that ten generations have been tested).
The convergence speed is defined as the gradient of the generational distance.

An optimisation is considered converged if the convergence speed $\leq \num{5e-4}$ for two consecutive generations and if the generational distance $\leq \num{5e-3}$.
The changes of the Pareto front are assumed to be minor after the convergence criterion is fulfilled.

\section{Results and discussion}
Here, we will analyse and compare the performance of the three selected parameterisation approaches for optimisation applications by the example of a multi-objective optimisation of the pitching kinematics of a flapping wing.
The pitching kinematics are parametrised by
\begin{inparaenum}[a)]
\item control points connected by a spline interpolation,
\item a finite Fourier series, and
\item a modal reconstruction of POD eigenmodes.
\end{inparaenum}
Each of the approaches has been tested with $6$, $12$, and $18$ parameters.
This lead to more than $\num{30 000}$ experimental iterations that have been executed over a period of several weeks.
We will first compare how well the different approaches cover the motion space, discuss the optimisation process and convergence, and finally compare the Pareto front solutions and the optimised kinematics.

\begin{figure*}
\centering
\includegraphics[width=\linewidth]{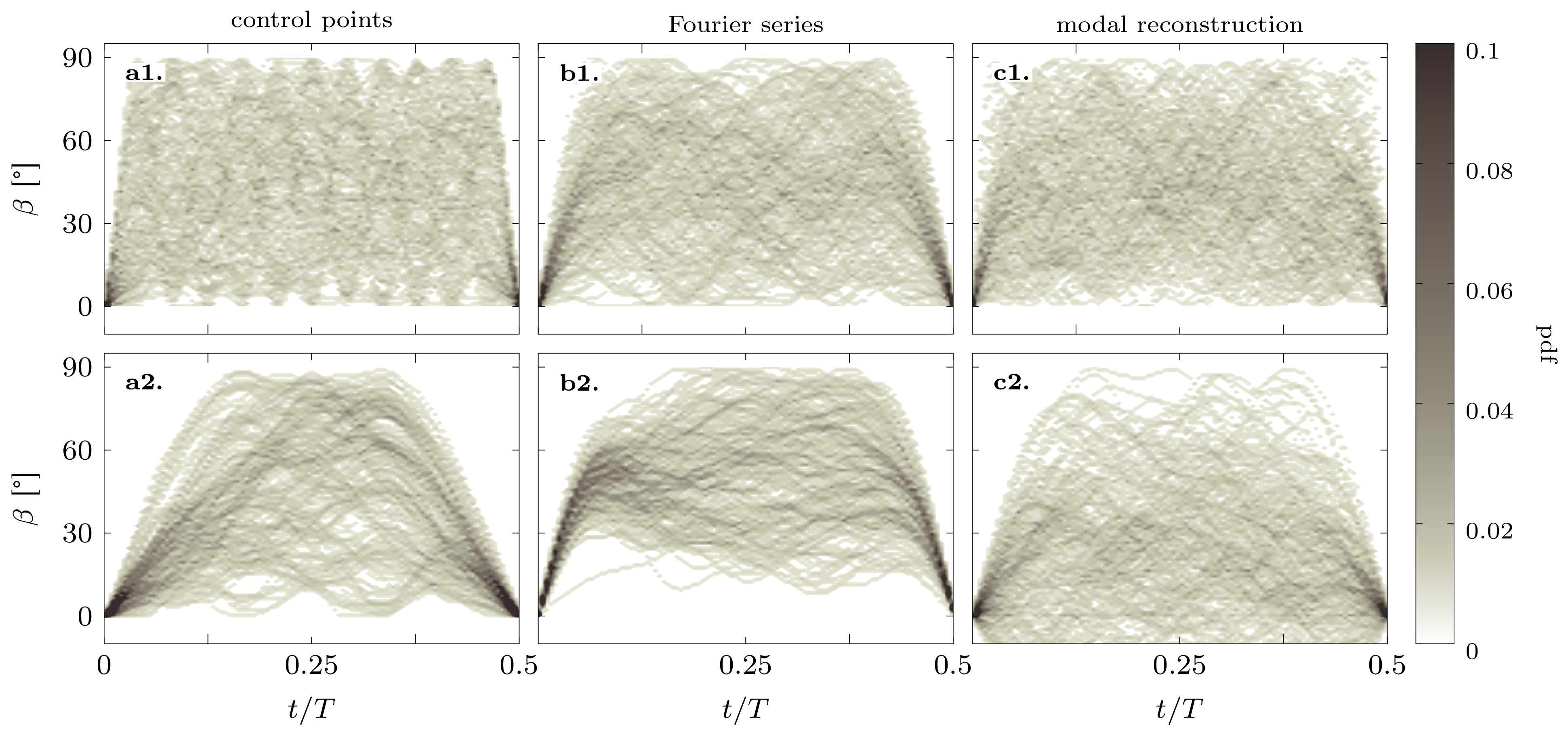}
\caption{Probability density map of a randomly selected initial population consisting of $200$ individuals for the different parameterisations with $12$ parameters without applying a phase-shift.
The pitching angle is bound by $\kindex{\beta}{min} = \SI{0}{\degree}$ and $\kindex{\beta}{max} = \SI{90}{\degree}$.
First row shows the coverage of the solutions space for the different parameterisation methods: a) control points connected by a spline interpolation, b) finite Fourier series, c)
modal reconstruction, without taking into account an acceleration constraint.
The second row shows the coverages when the acceleration constraint ($\kindex{\ddot{\beta}}{max} \leq \SI{1800}{\degree \per \second \squared}$) is applied.
The lower angular bound was lifted for the modal reconstruction when applying the acceleration constraint to allow for a set of $200$ executable individuals to be found in a reasonable amount of time.}
\label{fig:initial_pop}
\end{figure*}

\begin{figure*}
\includegraphics[width=\linewidth]{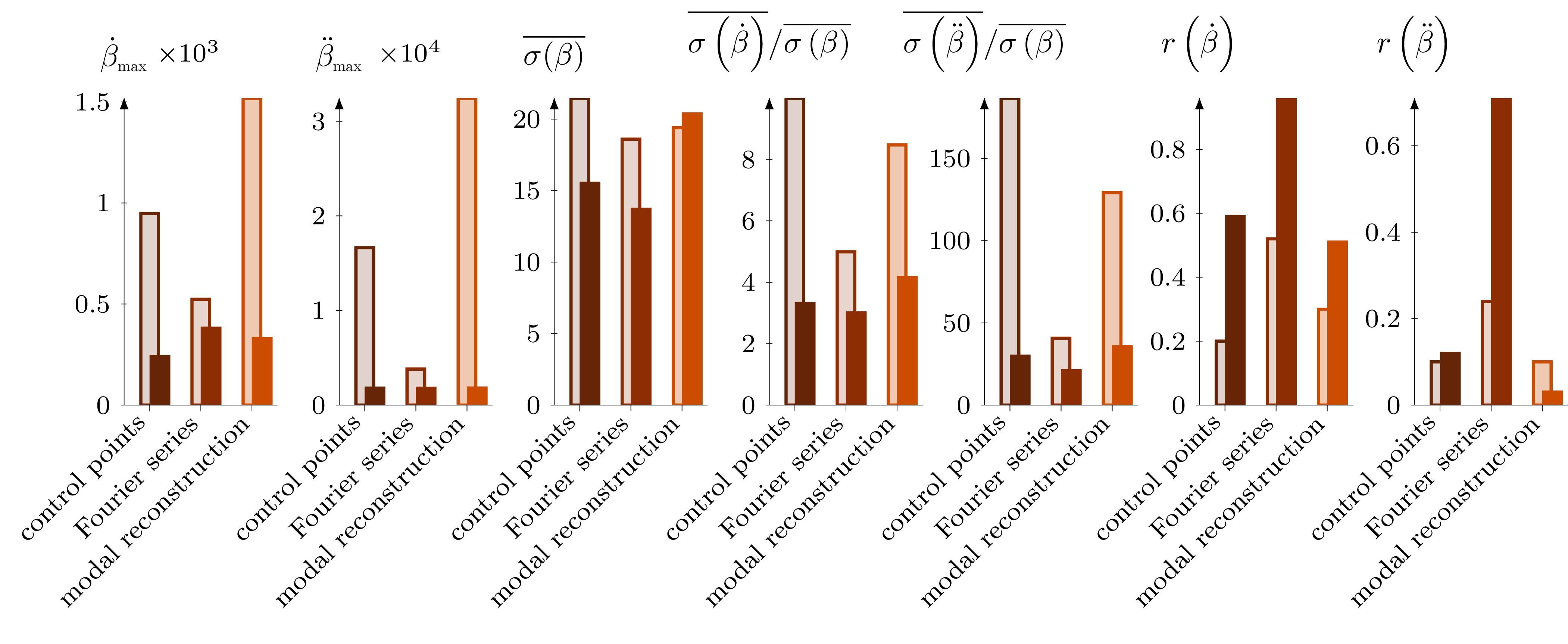}
\caption{Quantitative comparison of the characteristic properties of the initial population consisting of $200$ individuals for the different parametrised of the kinematics for $12$ parameters.
The brighter bars indicate values obtained without taking into account the acceleration constraint.
The darker bars include the influence of the acceleration constraint.
The diversity of the population is characterised by the maximum pitch rate $\kindex{\dot{\beta}}{max}$, maximum pitching acceleration $\kindex{\ddot{\beta}}{max}$,
the half-stroke-averaged standard deviation of the pitching angle $\overline{\sigma(\beta)}$, the half-stroke-averaged normalised standard deviation of the pitch rate $\overline{\sigma( {\dot{\beta}} )} / \overline{\sigma( {\beta} )} $ and the pitching acceleration $\overline{\sigma( {\ddot{\beta}} )} / \overline{\sigma( {\beta} )}$, and the correlation-coefficients for the pitch rate $ r ( \dot{ \beta } ) $
and pitching acceleration $r(\ddot{\beta})$.
}
\label{fig:pop_prop}
\end{figure*}

\subsection{Coverage of the kinematic solution space}
Suitable parameterisation approaches for optimisation studies should be able to represent a high variety of qualitative shapes and achieve good initial coverage of the solution space to not a priori exclude potential optimal solutions.
To quantify and compare the ability of the different approaches to cover the solution space, we randomly created an initial population consisting of $200$ individuals for the different parameterisations with $12$ parameters without any acceleration constraint and without applying the phase-shift.
The probability density maps for these initial populations of pitching kinematics are presented in the top row of \cref{fig:initial_pop} to give a visual impression of the coverage of the solution space.
The density maps for $p=18$ do not show significant differences and lead to the same conclusions as for $p=12$.
The splines connecting control points and the modal reconstruction functions cover almost all achievable angles in the first half-stroke (\cref{fig:initial_pop}a1,c1).
The Fourier-series have a slightly narrower band of admissible motions around stroke reversal with lower diversity (\cref{fig:initial_pop}b1).
The fixed phase-locations of the control points lead to the occurrence of regions in the density map with higher local probabilities indicating an inhomogeneous coverage and slight bias towards low angles at the control points.

A quantitative comparison is presented in \cref{fig:pop_prop}, where the brighter bars indicate the values without the application of the acceleration constraint.
The parameters we calculated to characterise the coverage of the solutions space and diversity of the parameterised motions kinematics include the maximum pitch rate $\kindex{\dot{\beta}}{max}$ and pitch acceleration $\kindex{\ddot{\beta}}{max}$, the half-stroke-averaged standard deviation of the pitching angle $\overline{\sigma(\beta)}$, the half-stroke-averaged normalised standard deviation of the pitch rate $\overline{\sigma( {\dot{\beta}} )} / \overline{\sigma( {\beta} )} $ and the pitch ac\-celera\-tion $\overline{\sigma( {\ddot{\beta}} )} / \overline{\sigma( {\beta} )} $.
The Fourier series-based kinematics have the lowest maximum pitch rate and pitch accelerations and a lower diversity based as indicated by lower values of the standard deviations of the pitch angle, rate, and acceleration compared to the control point approach and the modal reconstruction.
The highest gradients are measured among the modal reconstruction solutions if the acceleration constraint is not applied.
The control point kinematics exhibit the highest diversity closely followed by the modal reconstruction method if we do not apply the acceleration constraint.

If we apply the acceleration constraint to omit solutions that cannot or should not be executed by our experimental device, we obtain the probability density plots presented in the bottom row of \cref{fig:initial_pop}.
The quantitative measures after application of the acceleration constraint are depicted in \cref{fig:pop_prop} by the darker bars.
Here, the acceleration is limited to $\ddot{\beta} \leq \SI{1800}{\degree \per \second \squared}$.

The coverage achieved by the initial populations is now reduced for all approaches especially near the stroke reversal.
Motions with larger angles around stroke reversal are omitted due to the acceleration limit.
This affects the control point approach the most.
The splines connecting the control points tend to develop characteristic spikes when the distance between control points decreases which leads to accelerations above the allowed limit.
The initial population for Fourier series kinematics exhibits a small band of angles with very low diversity around stroke reversal and a poor coverage for smaller $\beta$.
This is caused by the different restriction of the parameter space for the Fourier series.
The best coverage is achieved for all kinematic functions at moderate angles between $t/T = 0.125-0.375$.
The modal reconstruction has the highest density at lower angles compared to the others (\cref{fig:initial_pop}c1).
Unfortunately, this increases the chance that a randomly selected individual does not meet the acceleration constraint and needs to be omitted and replaced when building he initial population.
To obtain a set of \num{200} executable individuals for the initial population within a reasonable amount of time, the lower angular constraint $\kindex{\beta}{min}$ is loosened.
This explains the non-zero probability for $\kindex{\beta}{min}<0$ in \cref{fig:initial_pop}c2.

All kinematic functions show a significant loss in diversity following the application of the acceleration constraint.
The modal reconstruction now outperforms the control point approach and the Fourier series in all measures.
Most of the diversity of the initial control point approach without constraint is created over the tightly spaced control points with a large spectrum of admissible angles.
This allows for a large variety of angles, velocities, and accelerations, but it also leads to higher number of omitted kinematics.
This could potentially be reduced by distributing the control points differently or by also considering their time coordinates as variable parameters.

Alternative measures for the diversity of the randomly selected individuals in an initial population are obtained by determining the correlation-coefficients for the pitch rate $r(\dot{\beta})$ and the pitch acceleration $r(\ddot{\beta})$.
They are calculated as the ensemble average correlation-coefficient $r$ for each individual with the rest of the population.
A higher diversity within a population now leads to a lower correlation coefficient $r$.
Overall, the diversity decreases upon application of the acceleration constraint as the correlation values increase.
The control point approach and the modal reconstruction show again a higher diversity than the population of Fourier series.

\begin{figure*}
\centering
\includegraphics[width=\linewidth]{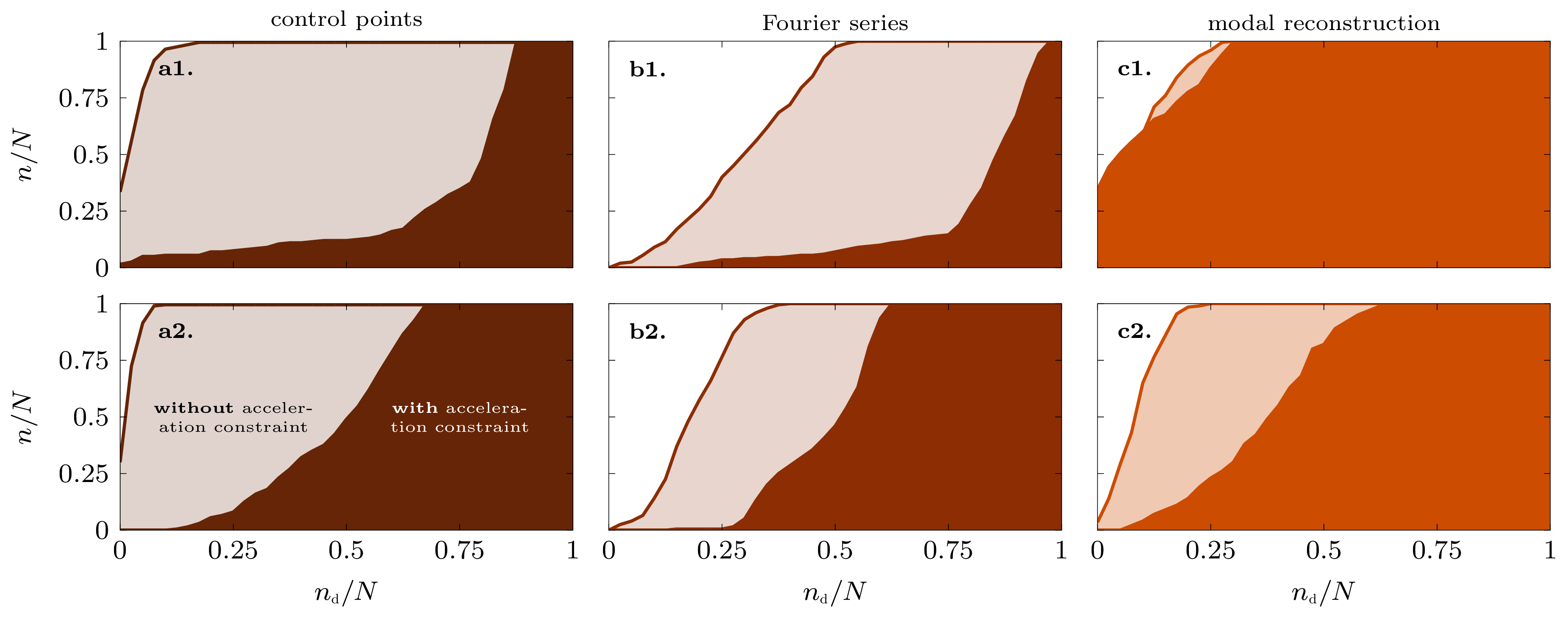}
\caption{Cumulative distribution of the relative node degree $\kindex{n}{d}/N$ in the similarly networks describing the initial population of $N=200$ individuals for the three parameterisation approaches.
Distributions corresponding to the networks based on $\kindex{R}{i,i}$ are in the top row, those based on $\kindex{D}{i,i}$ are in the bottom row.
The light and dark shading correspond respectively to randomly selected populations without and with the application of the acceleration constraint.
}
\label{fig:similarity}
\end{figure*}

To more quantitatively compare the similarity between different randomly generated kinematic, we borrow ideas here from a network topology approach proposed by \citet{zhang_complex_2006} and more recently used e.g. by \citet{tandon_condensation_2021} for comparing the similarity between trajectories.
The similarity between kinematics can be quantified using the maximum cross-correlation coefficient $\kindex{R}{i,j}$ for each pair of kinematics $\kindex{\beta}{i}$ and $\kindex{\beta}{j}$.
The maximum cross-correlation coefficient is defined as:
\begin{equation}
\kindex{R}{i,j} = \max\limits_{\tau=0...T} \frac{cov\left(\kindex{\beta}{i}(0:T),\kindex{\beta}{j}(\tau:(T+\tau))\right)}{\sqrt{var\left(\kindex{\beta}{i}(0:T)\right) var\left(\kindex{\beta}{j}(\tau:(T+\tau)\right)}}
\end{equation}
with $T$ the period, $cov(.)$ the covariance, and $var(.)$ the variance operators.
By taking the maximum over all possible phase shifts $\tau$, the resulting value only evaluates the shape similarity and is not influence by a phase shift, e.g. introduced by $\kindex{\Delta t}{0}$ here.
Another metric proposed by \citet{zhang_complex_2006} is the mean absolute error $\kindex{D}{i,j}$ between two curves, which is defined here as:
\begin{equation}
\kindex{D}{i,j} = \min\limits_{\tau=0...T} \displaystyle\frac{1}{T}\sum\limits_{t=0}^{T} \left\lvert\kindex{\beta}{i}(t)-\kindex{\beta}{j}(t+\tau)\right\rvert \quad.
\end{equation}
We normalised \kindex{D}{i,j} by the period $T$ and expressed the result in degree such that $\kindex{D}{i,j}$ can be interpreted as an average pitch angle amplitude difference between two kinematics.
Two identical curves have a maximum cross-correlation coefficient $\kindex{R}{i,j}=1$ and a minimal mean absolute error $\kindex{D}{i,j}=0$.
Both quantities evaluate different aspects, $\kindex{R}{i,j}$ provides a measure for the shape similarity between curves and $\kindex{D}{i,j}$ provides a measure for the amplitude variations between curves.
We use both here to evaluate the degree of similarity within a randomly generated population.

After calculating $\kindex{R}{i,j}$ and $\kindex{D}{i,j}$ between all curves with the initial population, we construct two similarity networks.
Each kinematic motion is represented by a node in the network.
Different nodes are connected to each other if the shape similarity measure $\kindex{R}{i,j}$ is higher than a predefined threshold \kindex{R}{th} or if the amplitude dissimilarity measure $\kindex{D}{i,j}$ is lower than a predefined threshold \kindex{D}{th}.
We then determine the node degree for each node, which corresponds to the number of nodes it is connected to.
Kinematics with a low node degree show similarity only to a low number of other kinematics in the population.
The cumulative distributions of the node degrees (\kindex{n}{d}) among the initial population for the three different parameterisation approaches with $p=12$ are presented in \cref{fig:similarity}.
The light and dark shading correspond respectively to randomly selected populations without and with the application of the acceleration constraint.
The more diverse a population of kinematics the more kinematics we expect with a low degree of connectivity and the larger the coloured area in \cref{fig:similarity}.
The distributions in the top row have been obtained for the networks based on the shape similarity measure $\kindex{R}{i,j}$ and indicate how diverse the population is in terms of the shape of the kinematics.
The distributions in the bottom row have been obtained for the networks based on the amplitude difference measure $\kindex{D}{i,j}$ and indicate how diverse the population is in terms of the amplitudes of the kinematics.


The unconstraint populations for the control point and the modal reconstruction have a high degree of diversity as the majority of the kinematics have a low node degree and show similarities to only a few other kinematics in the population.
The Fourier series performs slightly lower than the two others.
By construction, the Fourier series have an inherent similarity in shape and it is not entirely unexpected to find more nodes with a higher node degree for this approach.
Almost all cumulative distributions in \cref{fig:similarity} shift to higher node degrees when the acceleration constraint is applied, indicating that the constraint decreases the diversity both in terms of shapes and in terms of amplitudes.
The only exception is the shape diversity for the modal reconstruction which remains nearly unaffected by the constraint.
The spline population is the most diverse for the unconstrained case and the modal population exhibits the highest diversity for the constrained case.

We have also used these metrics to analyse the ability of the different approaches to create the insect-like kinematics and trapezoidal motions presented in \cref{fig:experimental_setup}.
All approaches except the modal reconstruction are capable to mimic the presented kinematics with a minimum correlation coefficient $R\geq 0.993$ and a maximum mean absolute error $D\leq\ang{1}$.
The modal reconstruction does equally well for most of the motions except for the trapezoid with steep edges which is approximated with a larger mean absolute error of $D=\ang{2.3}$.

Based on all visual and quantitative comparison presented here, the control point method yields the best coverage of the solution space and the largest diversity within an initial population if no acceleration constraint is applied.
The modal reconstruction method performs better than the control point method when the constraint is applied.
The Fourier series populations are least diverse and have the lowest coverage, but still perform sufficiently well for our optimisation application.

\subsection{Optimisation progress and convergence}

\begin{figure}
\centering
\includegraphics[width=\linewidth]{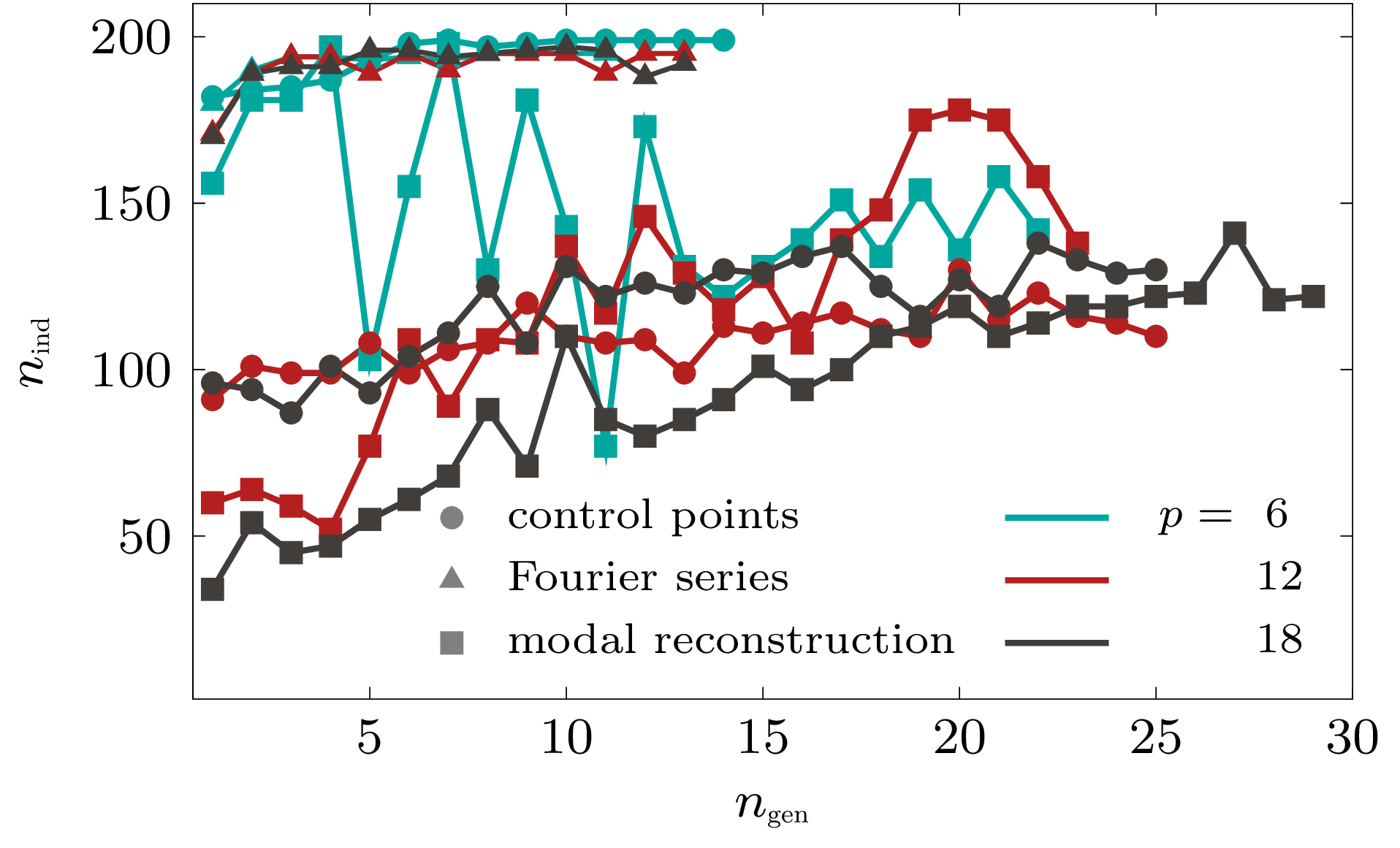}
\caption{Number of executable motions or individuals per generation for all optimisations with $6$, $12$, and $18$ parameters and the three parameterisation approaches.}
\label{fig:individuals_per_generation}
\end{figure}

Due the specific implementation of the genetic algorithm for our flapping wing optimisation, we can not replace individual kinematics that are omitted based on the acceleration constraint.
This is not a general limitation of the approach and can be overcome for future applications.
However, in the current study, the total number of the executable individuals in first generations is significantly lower than the nominal $200$ individuals due to the implementation (\cref{fig:individuals_per_generation}).
This is in particular the case for the control point and modal reconstruction methods with $12$ or $18$ parameters.
The Fourier series method is less affected and also the optimisation with a lower number of control points has populations with consistently close to \SI{85}{\percent} of executable individuals.
The number of executable individuals per generation increases for most of the optimisations but never reaches full population size with exception of the control point approach with $6$ parameters.
The control point approach with $12$ parameters and the modal reconstruction approach have particularly low numbers of executable individuals in the early generations, but the situation improves in the course of the optimisation.
The lower number of executable individuals in the populations decrease the diversity which hampers the search for the optimum solutions and delays convergence.

\begin{figure}
\centering
\includegraphics[width=\linewidth]{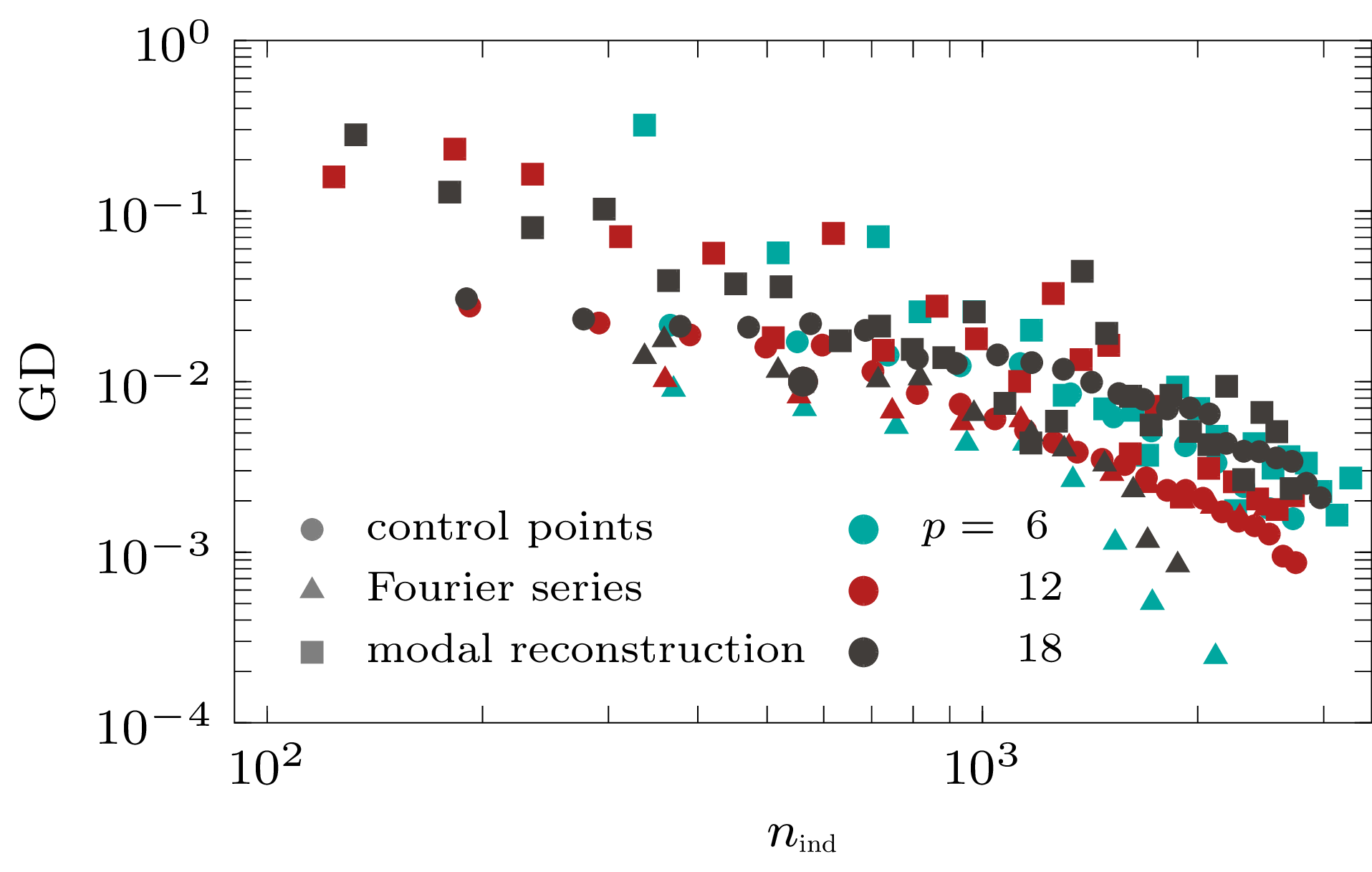}
\caption{Moving average over $10$ iterations for the generational distance (GD) versus the number of iterations.
Each iteration corresponds to the evaluation of one flapping motion.}
\label{fig:convergence}
\end{figure}

The convergence of the different approaches is analysed based on the generational distance introduced in \cref{eq:gd}.
The variation of the generational distance with the number of executed motions is presented in \cref{fig:convergence}.
Surprisingly, the parameter count is not the most important factor that influences the convergence behaviour.
The largest influence on the convergence is the parameterisation approach.
The optimisations using the same parameterisation form groups with similar convergence speed according to the generational distance metric.

The modal reconstructions exhibit the highest relative generational distance and convergence speed at the very beginning.
The modal reconstruction with $18$ parameters has the highest initial generational distance of \num{0.346} but also the modal approach with $12$ (GD=\num{0.129}) and $6$ parameters (GD=\num{0.125}) have initial values that are one magnitude larger than the values obtained for the other conditions.
The higher initial generational distance can be explained by a larger diversity of the initial population.
The convergence speed behaves proportionally to the generational distance such that all optimisations converge in comparable time.
The smallest relative generational distance and convergence speed is observed for the Fourier series.
The motions of the initial Fourier population are less diverse and closer to the final solution due to the different constriction of the parameter bounds.
The optimisation progress is slower because the differences between the motions are less pronounced and consequently harder to determine.
On the other side, higher diversity for the initial populations leads to higher convergence speed.

The total convergence time differs by \SI{40}{\percent} between the slowest (control point approach with $p=12$) and fastest (Fourier series approach with $p=6$) optimisation and the majority takes between $\num{2486}$ (Fourier series approach with $p=12$) and $\num{2742}$ (control point approach with $p=12$) iterations (\cref{tab:convergence}).
The optimisations are not deterministic and uncertainties are introduced due to random initialisation, recombination and mutation.
The differences in convergence time are small when considering that the computational complexity of the optimisations increases exponentially with the parameter count.
The Fourier series optimisations tend to reach convergence slightly faster than the modal and control point approaches.


\begin{table}
\centering
\caption{Number of iterations and generations required until convergence for the different parameterisation approaches.
Convergence is reached if the convergence speed $\leq \num{5e-4}$ and generational distance $ \leq \num{5e-3}$ for two consecutive generations.}
\begin{tabular}{@{}llll@{}}
\toprule
parameterisation & $p$ & iterations & generations\\
\toprule
control points & 6 & $2718$ & $14$\\
& 12 & $2742^\ast$ & $25^\ast$ \\
& 18 & $2968$ & $25$ \\
Fourier series & 6 & $2118$ & $11$\\
& 12 & $2486$ & $13$\\
& 18 & $2491$ & $13$\\
modal reconstruction & 6 & $2683$ & $18$ \\
& 12 & $2721$ & $23$ \\
& 18 & $2702$ & $29$\\
\toprule
\multicolumn{4}{l}{\scriptsize
\parbox{0.85\linewidth}{
$^{*}$ The convergence for this case is determined manually by observing the evolution of the Pareto kinematics.}}
\end{tabular}
\label{tab:convergence}
\end{table}{}

\subsection{Pareto fronts and optimal kinematics}

\begin{figure*}
\centering
\includegraphics[width=\linewidth]{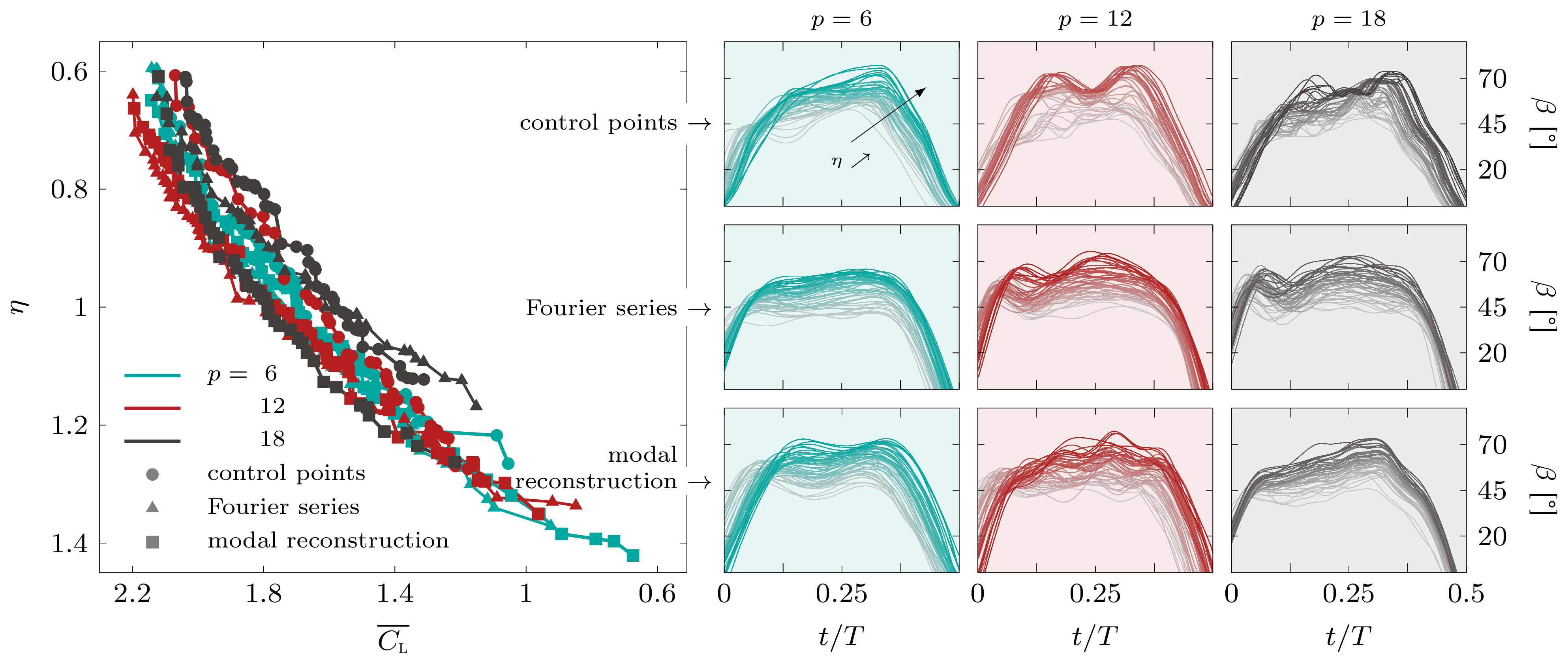}
\caption{Global Pareto front for all kinematic function optimisations for the last generation.
Kinematics corresponding to the final global Pareto front solutions for the different kinematic functions.}
\label{fig:pareto_front}
\end{figure*}

The resulting Pareto fronts for all $9$ optimisations are presented in \cref{fig:pareto_front}.
The colours indicate the number of optimisation parameters and the symbols indicate the parameterisation approach.
The different Pareto fronts have the same shape but there is a shift between them and they reach slightly different extreme values.
The modal reconstruction approach with $p=6$ finds the most efficient solutions (up to $\eta\approx1.4$) and the Fourier series approach with $p=12$ finds the solutions that generate the highest stroke average lift (up to $\kindex{\overline{C}}{L}\approx 2.2$).
Overall, the optimisations with the highest number of parameters are most limited in finding highly efficient solutions.
The higher number of parameters might lead to more complex kinematics that require more power to achieve the same performance in lift.
For our current example, $p=12$ seems to provide the best balance across the different parameterisation approaches.
The additional complexity provided by $12$ optimisation parameters compared to the $p=6$ optimisations is beneficial in improving the aerodynamic performance without increasing the power requirements and penalising the efficiency.
Increasing the parameter count leads to slower convergence and kinematics with higher harmonics variations.
To aid the genetic algorithm to fine tune the parameters, we have conducted an additional optimisation where we have taken the results for $p=6$ as a start point for a $p=12$ optimisation using the Fourier series parameterisation approach.
The results did not lead to significant differences with respect to the regular $p=12$ optimisation and are not presented here.
Overall, the differences between Pareto fronts are of the same order of magnitude as the inherent experimental variations.
The influence of the number of parameters seems to have a stronger influence than the parameterisation approach.

In general, the performance of the three parameterisations is comparable and we can not identify a clear preference of any one of them.
The optimal kinematics that make up the Pareto front are included in \cref{fig:pareto_front}.
The columns with the same background colour correspond to optimisations with the same number of parameters.
The rows correspond to different parameterisation approaches.
The colour of the $\beta(t/T)$ curves for one half stroke varies from grey to green, red, or black with increasing efficiency.
All Pareto kinematics show a continuous transition from high lift motions to more efficient ones, which indicates that the solutions are converged and we obtained a well developed Pareto front.
For the same number of parameters, the Pareto kinematics show similar characteristic features for the three parameterisation approaches but also display some subtle differences that cause the shifts in the Pareto front.

A clear common feature is the preference for advanced rotation for most Pareto kinematics.
Advanced rotation means that the wing has rotated past its horizontal orientation ($\beta=0$) before the end of the stroke motion.
By doing so, it starts the next stroke with a lower and more favourable angle of attack.
A positive influence of the advancement of the rotation for lift production has been observed before on insects and their robotics counterparts \cite{Sane.2001, Sun.2002, Krishna:2019cb}.
The rotational advancement generates a higher stroke average lift but comes at a cost and there is a continuous decrease of the phase shift towards less advanced rotation with increasing efficiency in all our optimisations.
The pitch rate at the end of the stroke is strikingly similar for all cases.
The variation in phase shift between the highest lift generating and most efficient kinematics tends to decrease for optimisations with fewer parameters.
This suggests that lift increase can be generated at lower cost by a more complex pitching kinematics during the entire stroke instead of relying of the advanced rotation (\cref{fig:pareto_front}).

Another common feature is the increase in the maximum pitch angle with increasing efficiency.
A larger pitch angle corresponds to a lower angle of attack with leads to lower drag and more efficient kinematics.
The values of the maximum pitch angles are approximately the same for all parameterisations and parameter counts except for the Fourier series, which reach slightly lower maximum values of $\beta$.

The main differences between the Pareto optimal kinematics of the different cases are the number of local maxima and their timing.
The control point and modal reconstruction approaches with $p=6$ show two local maxima around $t/T = 0.125$ and $t/T = 0.375$ (\cref{fig:pareto_front}).
These two maxima are less pronounced or even absent in the Fourier series kinematics which look more like a smooth trapezoidal profile.
The two local maxima become more pronounced for the control point approach with $p=12$ and only the second one around $t/T = 0.375$ remains present when $p$ is increased to $18$ for this approach.
The modal reconstructions for $p=18$ are also characterised by a single peak that is located around $t/T = 0.32$.
The Fourier series with $p=12$ and $p=18$ have a distinct bump at the beginning of the stroke, around $t/T=0.1$.
Overall, the complexity of the motions increases with the parameter count and different shapes and kinematics can lead to similar Pareto fronts depending on the choice of the parameterisation and the parameter count.
This is due to the complex relationship between the flow development and the growth of the leading edge vortex for flapping wings motions \cite{Bhat.2020,gehrke_phenomenology_2021}.

In previous work, we identified the shear layer velocity $\kindex{u}{s}$ at the leading edge as a characteristic parameter to scale the force response of generalised flapping wing pitching kinematics \citep{gehrke_phenomenology_2021}.
The shear layer velocity act as a proxy for the feeding rate of the leading edge vortex.
Higher angles of attack or lower pitching angles lead to a higher shear layer velocity $\kindex{u}{s}$ which in turn leads to a faster growth of the leading edge vortex.
This is the common feature of high lift generating kinematics.
The stronger leading edge vortex leads to higher lift, but reaches its maximum size and circulation earlier in the cycle.
The maximum in lift and leading edge circulation is typically reached around mid-stroke for high lift kinematics.
After the vortex reaches its maximum circulation, it lifts off of the wing which leads to a drop in lift in second half of the stroke.

Efficient motions exhibit smaller angles of attack with pitching angle maxima around $\kindex{\beta}{max}=\SI{70}{\degree}$ (\cref{fig:pareto_front}).
They create a smaller leading edge vortex which grows continuously during the majority of the stroke and promote drag-minimised lift creation.
The power and lift coefficients are less dominated by the stroke motion as the angle of attack is smaller causing the shear layer velocity \kindex{u}{s} to be relatively constant during mid-stroke.

\section{Conclusion}

In this study, we presented three different approaches to parameterise motion kinematics, demonstrated their application, and evaluated their performance by the example of an experimental optimisation of the pitching kinematics of a robotic flapping wing in hover.
The pitch angle kinematics in our application are described by
\begin{inparaenum}[a)]
\item control points that are connected by a fifth-order spline interpolation,
\item a finite Fourier series, and
\item a linear combination of modes determined by a modal decomposition of kinematics created by a random walk.
\end{inparaenum}
Each of the parameterisation approaches is implemented with three different parameter counts: $6$, $12$ and $18$, which leads to a total of nine optimisations and more than $\num{30 000}$ experimental iterations conducted over a period of several weeks.
The performance of the different approaches has been analysed by comparing the diversity of the solutions and coverage of the motion space, the optimisation process and convergence, and the Pareto front solutions and the optimised kinematics.

The coverage of the solution space was qualitatively evaluated based on a probability density map of the random initial population of \num{200} individuals.
For a more quantitative comparison of the coverage and the diversity of the solutions, we introduced a number of diversity measures including the maximum pitch angle, rate, and acceleration, their standard deviations, and correlation coefficients for the pitch and acceleration.
To further quantify the diversity between the kinematics within the initial populations, we calculated similarity networks based on a mutual shape similarity measure and an amplitude error.
The cumulative distribution of the node degree in a population is an intuitive quantitative measure to compare the diversity of the kinematics based on their shape and amplitude.
Based on all qualitative and quantitative measures, the control point kinematics exhibit the highest diversity closely followed by the modal reconstruction method if no experimental constraints are in place.
The Fourier series-based kinematics have the lowest maximum pitch rate, pitch acceleration, and a lower diversity compared to the control point and the modal reconstruction approach.

In most experimental applications, the theoretical parameter space cannot be fully explored due to mechanical, electrical, or other constraints for example related to the measurement equipment.
In our case, the main limitation is imposed by the sensitivity and measurement range of the load cell and the performance envelope of the motor that controls the pitching motion.
These limitations lead to a pitch acceleration constraint that significantly reduces the coverage and the solution diversity for all three parameterisation approaches.
The practical implementation of an acceleration constraint is easier for the Fourier series approach than for the two other approaches.
The modal reconstruction parameterisation performs best in the various diversity measures, followed by the control point and the Fourier series approaches when the acceleration constraint is in place.

The generational distance criterion is implemented to monitor the progress of the optimisation and as a convergence measure to end the optimisation.
All optimisations converged in comparable time which did not depend on the parameter count.
In the presented optimisation study, it is desirable to have fewer parameters as more parameters increase complexity without a clear aerodynamic benefit.
The subtle changes in the kinematics lead to variations in the loads that are of the same order as the inherent experimental fluctuation of the system.
This can slow down the convergence of the genetic algorithm especially in later stages when only minor aerodynamic improvements are achieved.

The resulting Pareto fronts for the different optimisations are similar in shape and in the range of values of the two objective functions they cover.
The parameter count has a stronger influence on the final results than the parameterisation approach.
This is encouraging for the robustness of the three parameterisation approaches presented here.
The optimal kinematics corresponding to the Pareto fronts have slightly different shapes depending on the parameter count and the parameterisation approach selected but they yield comparable fitness values.
The most prominent common features of the kinematics are their shared preference for advanced rotation and the increase in the maximum pitch angle with increasing efficiency.
The main differences between the optimal kinematics of the different cases are the number of local maxima and their timing.

\begin{table*}[tb!]
\centering
\caption{Overview of the key characteristics for the different parameterisations.}
\begin{tabular}{l@{}cccc@{}}
\toprule
& Complexity & Handling  & Interpretability & Applications \\
&  & of constraints & of parameters & \\
\toprule
control points & - & + & + & intermittent kinematics, e.g. \\
& & & & burst and coast swimming \\
Fourier series & + & $\circ$ & - & vertical axis wind turbines, \\
& & & & flapping foil energy harvester \\
modal reconstruction & $\circ$ & -  & $\circ$ & complex animal-like \\
& & & & locomotion \\
\toprule
\end{tabular}
\label{tab:conclusion}
\end{table*}

The goal of this paper was to present three different methods to parameterise complex motion kinematics for optimisation studies and to compare their performance on an experimental flapping wing system.
Each optimisation application has different objective functions, constraints, and specific challenges.
There is no one solution that fits all and the different approaches to parameterise the solutions have their specific advantages and disadvantages outlined in~\cref{tab:conclusion}.
The Fourier series approach is the easiest method to implement but the introduction of parameter constraints and the parameter bounds cannot be as directly contained as for the control point approach.
Nevertheless, the Fourier series has an analytical definition and continuous high-order derivatives which make it a promising parameterisation for structurally sensitive applications like vertical-axis wind turbines or flapping-foil energy harvesters.
By definition, the control points are more intuitive to constrain and their parameter values can be interpreted directly in terms of their influence on the temporal evolution and amplitude of the kinematics.
With the ability to create sharp and zero gradient curves, the control point method is most suitable for intermittent kinematics like burst and coast swimming of fish- and squid-like devices, and flapping wing flight.
The modal reconstruction is extremely versatile and adaptable.
The initial family of solutions can be generated by a random walk method as presented here, but can also be a family of kinematics measured in nature or in the lab.
Building a parameterisation upon an existing library of kinematics can be an extremely powerful tool.
Complex animal-like locomotions can be described by only a few modes which would require many parameters being constructed by analytical kinematic definitions.

Depending on the individual complexity of the mode shapes, it is almost impossible to restrict parameters bounds to meet experimental constraints.
Instead, motions have to be omitted after generation if they do not meet constrains such as start and end position and acceleration.
This can severely hamper the creation of initial populations for a genetic algorithm optimisation, for example.
In the presented application of an experimental optimisation of a robotic flapping wing mechanism, all three approaches yielded similar results in a comparable amount of time.
Other applications might have different challenges, but we expect that the main characteristics of the presented approaches are valid for different systems and that the choice of the most suitable parameterisation approach can be guided by the properties listed in \cref{tab:conclusion}.


\section*{Acknowledgments}
\addcontentsline{toc}{section}{Acknowledgments} %

The authors thank Guillaume de Guyon for the help in implementing the fifth-degree spline kinematic function.
Funding has been provided by the Swiss National Science Foundation under grant number 200021\_175792.

\bibliography{param4opt}

\begin{thebibliography}{27}
\providecommand{\natexlab}[1]{#1}
\providecommand{\url}[1]{{#1}}
\providecommand{\urlprefix}{URL }
\expandafter\ifx\csname urlstyle\endcsname\relax
  \providecommand{\doi}[1]{DOI~\discretionary{}{}{}#1}\else
  \providecommand{\doi}{DOI~\discretionary{}{}{}\begingroup
  \urlstyle{rm}\Url}\fi
\providecommand{\eprint}[2][]{\url{#2}}

\bibitem[{Ansari et~al(2010)Ansari, Phillips, Stabler, Wilkins, Żbikowski, and
  Knowles}]{taylor_experimental_2010}
Ansari SA, Phillips N, Stabler G, Wilkins PC, Żbikowski R, Knowles K (2010)
  Experimental investigation of some aspects of insect-like flapping flight
  aerodynamics for application to micro air vehicles. In: Taylor GK,
  Triantafyllou MS, Tropea C (eds) Animal Locomotion, Springer Berlin
  Heidelberg, pp 215--236

\bibitem[{Audet et~al(2021)Audet, Bigeon, Cartier, {Le Digabel}, and
  Salomon}]{audet_performance_2020}
Audet C, Bigeon J, Cartier D, {Le Digabel} S, Salomon L (2021) Performance
  indicators in multiobjective optimization. European Journal of Operational
  Research 292(2):397--422, \doi{https://doi.org/10.1016/j.ejor.2020.11.016}

\bibitem[{Bayiz et~al(2018)Bayiz, Ghanaatpishe, Fathy, and
  Cheng}]{bayiz_hovering_2018}
Bayiz Y, Ghanaatpishe M, Fathy H, Cheng B (2018) Hovering efficiency comparison
  of rotary and flapping flight for rigid rectangular wings via dimensionless
  multi-objective optimization. Bioinspiration \& Biomimetics 13(4):046,002

\bibitem[{Berman and Wang(2007)}]{berman_energy-minimizing_2007}
Berman GJ, Wang ZJ (2007) Energy-minimizing kinematics in hovering insect
  flight. Journal of Fluid Mechanics 582:153--168

\bibitem[{Bhat et~al(2020)Bhat, Zhao, Sheridan, Hourigan, and
  Thompson}]{Bhat.2020}
Bhat SS, Zhao J, Sheridan J, Hourigan K, Thompson MC (2020) {Effects of
  flapping-motion profiles on insect-wing aerodynamics}. Journal of Fluid
  Mechanics 884:A8, \doi{10.1017/jfm.2019.929}

\bibitem[{Chipperfield et~al(1995)Chipperfield, Fleming, Pohlheim, and
  Fonseca}]{matlab_global_2020}
Chipperfield A, Fleming PJ, Pohlheim H, Fonseca CM (1995) The {MATLAB}
  {Genetic} {Algorithm} {Toolbox}. IEE, vol 1995, pp 10--10,
  \doi{10.1049/ic:19950061},
  \urlprefix\url{http://digital-library.theiet.org/content/conferences/10.1049/ic_19950061}

\bibitem[{Floreano and Wood(2015)}]{floreano_science_2015}
Floreano D, Wood RJ (2015) Science, technology and the future of small
  autonomous drones. Nature 521(7553):460--466

\bibitem[{Gehrke and Mulleners(2021)}]{gehrke_phenomenology_2021}
Gehrke A, Mulleners K (2021) Phenomenology and scaling of optimal flapping wing
  kinematics. Bioinspiration \& Biomimetics 16(2):026,016,
  \doi{10.1088/1748-3190/abd012}

\bibitem[{Gehrke et~al(2018)Gehrke, Guyon-Crozier, and
  Mulleners}]{gehrke_genetic_2018}
Gehrke A, Guyon-Crozier G, Mulleners K (2018) Genetic {Algorithm} {Based}
  {Optimization} of {Wing} {Rotation} in {Hover}. Fluids 3(3):59,
  \doi{10.3390/fluids3030059}

\bibitem[{Hamming(1986)}]{hamming_numerical_1986}
Hamming RW (1986) Numerical methods for scientists and engineers, 2nd edn.
  Dover

\bibitem[{Hawkes and Lentink(2016)}]{hawkes_fruit_2016}
Hawkes EW, Lentink D (2016) Fruit fly scale robots can hover longer with
  flapping wings than with spinning wings. Journal of The Royal Society
  Interface 13(123):20160,730

\bibitem[{Krishna et~al(2018)Krishna, Green, and
  Mulleners}]{krishna_flowfield_2018}
Krishna S, Green MA, Mulleners K (2018) Flowfield and {Force} {Evolution} for a
  {Symmetric} {Hovering} {Flat}-{Plate} {Wing}. AIAA Journal 56(4):1360--1371,
  \doi{10.2514/1.J056468}

\bibitem[{Krishna et~al(2019)Krishna, Green, and Mulleners}]{Krishna:2019cb}
Krishna S, Green MA, Mulleners K (2019) {Effect of pitch on the flow behavior
  around a hovering wing}. Experiments in Fluids 60(5):60--86,
  \doi{10.1007/s00348-019-2732-3}

\bibitem[{Lamont and Veldhuizen(1999)}]{veldhuizen_multiobjective_1999}
Lamont G, Veldhuizen DV (1999) Multiobjective evolutionary algorithms:
  classifications, analyses, and new innovations

\bibitem[{Liu and Aono(2009)}]{liu_size_2009}
Liu H, Aono H (2009) Size effects on insect hovering aerodynamics: an
  integrated computational study. Bioinspiration \& Biomimetics 4(1):015,002

\bibitem[{Manar et~al(2014)Manar, Medina, and Jones}]{manar_tip_2014}
Manar F, Medina A, Jones AR (2014) Tip vortex structure and aerodynamic loading
  on rotating wings in confined spaces. Experiments in Fluids 55(9),
  \doi{10.1007/s00348-014-1815-4}

\bibitem[{Mandre et~al(2019)Mandre, Breuer, and
  Miller}]{mandre_optimizing_2019}
Mandre S, Breuer KS, Miller MJ (2019) Optimizing the periodic motion of a foil
  at high-reynolds number. In: 9th International Symposium on Adaptive Motion
  of Animals and Machines (AMAM 2019), EPFL, Lausanne, Switzerland,
  \doi{10.5075/epfl-BIOROB-AMAM2019-63}

\bibitem[{Martin and Gharib(2019)}]{martin_experimental_2018}
Martin N, Gharib M (2019) Experimental trajectory optimization of a flapping
  fin propulsor using an evolutionary strategy. Bioinspiration \& Biomimetics
  14(1):016,010

\bibitem[{Sane and Dickinson(2001)}]{Sane.2001}
Sane SP, Dickinson MH (2001) {The control of flight force by a flapping wing:
  lift and drag production}. Journal of Experimental Biology
  204(15):2607--2626, \doi{10.1242/jeb.204.15.2607}

\bibitem[{Shyy(2011)}]{shyy_aerodynamics_2011}
Shyy W (2011) Aerodynamics of low Reynolds number flyers. Cambridge Univ. Pr.,
  {OCLC}: 730008577

\bibitem[{Shyy et~al(1999)Shyy, Berg, and Ljungqvist}]{shyy_flapping_1999}
Shyy W, Berg M, Ljungqvist D (1999) Flapping and flexible wings for biological
  and micro air vehicles. Progress in Aerospace Sciences 35(5):455--505

\bibitem[{Shyy et~al(2010)Shyy, Aono, Chimakurthi, Trizila, Kang, Cesnik, and
  Liu}]{shyy_recent_2010}
Shyy W, Aono H, Chimakurthi S, Trizila P, Kang CK, Cesnik C, Liu H (2010)
  Recent progress in flapping wing aerodynamics and aeroelasticity. Progress in
  Aerospace Sciences 46(7):284--327

\bibitem[{Sum~Wu et~al(2019)Sum~Wu, Nowak, and Breuer}]{sum_wu_scaling_2019}
Sum~Wu K, Nowak J, Breuer KS (2019) Scaling of the performance of
  insect-inspired passive-pitching flapping wings. Journal of The Royal Society
  Interface 16(161):20190,609, \doi{10.1098/rsif.2019.0609},
  \urlprefix\url{https://royalsocietypublishing.org/doi/10.1098/rsif.2019.0609}

\bibitem[{Sun and Tang(2002)}]{Sun.2002}
Sun M, Tang J (2002) {Unsteady aerodynamic force generation by a model fruit
  fly wing in flapping motion}. Journal of Experimental Biology 205(1):55--70,
  \doi{10.1242/jeb.205.1.55}

\bibitem[{Tandon and Sujith(2021)}]{tandon_condensation_2021}
Tandon S, Sujith RI (2021) Condensation in the phase space and network topology
  during transition from chaos to order in turbulent thermoacoustic systems.
  Chaos 31(4):043,126, \doi{10.1063/5.0039229},
  \urlprefix\url{https://aip.scitation.org/doi/10.1063/5.0039229}

\bibitem[{Tuncer and Kaya(2005)}]{tuncer_optimization_2005}
Tuncer IH, Kaya M (2005) Optimization of flapping airfoils for maximum thrust
  and propulsive efficiency. {AIAA} Journal 43(11):2329--2336

\bibitem[{Zhang and Small(2006)}]{zhang_complex_2006}
Zhang J, Small M (2006) Complex network from pseudoperiodic time series:
  Topology versus dynamics. Phys Rev Lett 96(23):238,701,
  \doi{10.1103/PhysRevLett.96.238701},
  \urlprefix\url{https://link.aps.org/doi/10.1103/PhysRevLett.96.238701}

\end{thebibliography}


\end{document}